\documentclass [useAMS,usenatbib]{mn2e}
\usepackage[dvips]{color,epsfig,rotating}
\usepackage{graphics}

\begin{document}

\title[Quasar Radio-Loudness and Core Ellipticals]{Quasar Radio-Loudness and the Elliptical Core Problem}

\author[T. S. Hamilton]{Timothy S. Hamilton$^{1}$\thanks{E-mail:
thamilton@shawnee.edu}\\
$^{1}$Department of Natural Sciences, Shawnee State University, 940 2nd St., Portsmouth, Ohio 45662, U.S.A.}

\date{}

\pagerange{\pageref{firstpage}--\pageref{lastpage}} \pubyear{2009}

\maketitle

\begin{abstract}

The dichotomy between radio-loud and radio-quiet QSOs is not simply one of host morphology.  While spiral galaxies almost exclusively host radio-quiet QSOs, ellipticals can host either radio-louds or radio-quiets.  We find that a combination of accretion rate and host scale determines which type of QSO a given elliptical galaxy will host.  QSOs with high x-ray luminosities (above $10^{44.5}$ erg s$^{-1}$ at 0.5 keV) are mostly radio-loud.  But those with low luminosities divide fairly neatly in size (measured by the half-light radius, $r_\mathrm{e}$).  Those larger than about 10 kpc are radio-loud, while smaller ones are radio-quiet.  It has recently been found that core and coreless  ellipticals are also divided near this limit.  This implies that for low-luminosity QSOs, radio-louds are found in core ellipticals, while radio-quiets are in coreless ellipticals and spirals.  This segregation also shows up strongly for low-redshift objects, and in general, there is a loss over time of coreless, radio-loud QSOs.  Since the presence or absence of a core may be tied to the galactic merger history, we have an evolutionary explanation for the differences between radio-loud and radio-quiet QSOs.
\end{abstract}


\section{Introduction}

\subsection{Studies of the Centres of Nearby Galaxies}

There has been considerable observational work in recent years on the kinematics and light profiles of the centres of nearby galaxies.  Several teams have worked on relationships with galaxies' central black holes.  Magorrian et al.~(1998) find that central black hole mass is proportional to bulge mass, while Ferrarese \& Merritt~(2000) and Gebhardt et al.~(2000) show it is related to central velocity dispersion.  Other studies have looked at central light profiles, such as Crane et al.~(1993), Faber et al.~(1997), Trujillo et al.~(2004), and Kormendy et al.~(2009).  These later two studies represent different approaches to modeling the inner light profiles, with Kormendy et al.~(2009) advocating a distinction between `core' galaxies (with flattened profiles at their centres) and `coreless' galaxies (with extra light at their centres compared to a S\'{e}rsic profile), and Trujillo et al.~(2004) advocating a distinction between `S\'{e}rsic' galaxies (well-fit by a S\'{e}rsic profile, including most of the coreless galaxies of Kormendy et al.~2009) and `core' galaxies (with shallow central profiles, but modeled differently than the other team's).

The imaging of QSO hosts, though making great strides with the Hubble Space Telescope (Bahcall, Kirhakos \& Saxe~1997; McLure et al.~1999; Hamilton, Casertano \& Turnshek~2002) and ground-based adaptive optics (Guyon, Sanders \& Stockton~2006), is mostly limited to morphology on a larger scale.  Many of these detailed observations cannot be done on QSOs, because of their distance and the glare from the bright nucleus (the black hole/bulge mass relation being an exception).  Thus we would like to associate known QSO properties with those more detailed characterizations of nearby galaxies.  

\subsection{Quasar Radio Loudness}

One of these known QSO properties is radio-loudness.  The major division of quasars into radio-loud and radio-quiet classes has been recognized for many years.  Early on, analogies to radio galaxies and Seyferts suggested that radio-loud QSOs would be found in elliptical hosts, and radio-quiets would be found in spirals.  Falcke, Sherwood \& Patnaik~(1996) challenged the simple division into radio-loud and radio-quiet with the suggestion of a radio-intermediate category.  Bahcall et al.~(1997) then found with the refurbished Hubble Space Telescope that while radio-quiet QSOs were, indeed, found in spiral (and interacting) hosts, radio-louds could be found in both ellipticals and spirals (and interacting hosts, again).  Investigations into the origins of radio emission in radio-loud QSOs have been done by Wang, Ho \& Staubert~(2003), and in radio-quiet QSOs (which can emit a detectable radio signal) by Laor \& Behar~(2008).  

\subsection{Fundamental Planes and Classification Tools}

Since the discovery of the fundamental plane of elliptical galaxies (Djorgovski \& Davis~1987; Dressler et al.~1987), others have found `fundamental planes' (two-dimensional distributions within higher-dimensional parameter spaces) in other relationships, such as that of black hole activity (Merloni, Heinz \& di Matteo~2003), that of x-ray gas in elliptical galaxies (Diehl \& Statler~2005), and even one for gamma-ray bursts (Tsutsui et al.~2009).  In our previous paper (Hamilton, Casertano \& Turnshek~2008), we identify a fundamental plane for QSOs that relates nuclear luminosity to the size and effective surface magnitude of the host.  This QSO fundamental plane is derived from a Principal Components Analysis (PCA), a multidimensional least-squares fitting technique. 

But because PCA identifies the sample's principal axes, the directions of maximum variance, this technique can also be used to identify different classes of objects.  When plotting the sample in the principal axis space, the classes may form separate clusters.  For example, this has been done for spiral galaxies by Whitmore~(1984), the stellar populations of early-type galaxies by Trager et al.~(2000), and QSO spectra by Boroson \& Green~(1992) and Yip et al.~(2004).  So a natural follow-up to our work on the QSO fundamental plane is to see if it can be used as a classification tool.

\subsection{Outline}

In this paper, we use the QSO fundamental plane as a classification tool (\S\ref{sec:qfp}) for radio-loud and radio-quiet quasars.  Our sample is chosen to probe a broad range of nuclear and host properties.  It contains nearly equal numbers of radio-loud and radio-quiet objects, and over three quarters of the hosts are elliptical galaxies, which are of particular interest because they can have either type of QSO.  We connect these fundamental plane results to large-scale physical properties of the QSOs, namely size and nuclear luminosity (\S\ref{sec:kormendy}), show how radio-loudness is associated with a combination of these properties, and examine their evolution with redshift.  Next, we propose an association between the distribution of QSO radio-loudness and the presence or absence of a `core' in the host (\S\ref{sec:association}).  In the Discussion (\S\ref{sec:discussion}), we review the results in the context of nuclear accretion mechanisms, and we discuss possible biases and evolutionary processes.

\section{The QSO Fundamental Plane}\label{sec:qfp}

This study uses the Hamilton et al.~(2008) sample of 42 low-redshift ($0.06 < z < 0.46$) QSOs with total nuclear plus
host light of $M_V \leq -23$ mag, taken from the Hubble Space Telescope (HST) archives.  All were observed with the Wide Field Planetary Camera 2 in broad-band optical filters.  These form a subset of our larger study of 70 QSOs, described by Hamilton et al.~(2002), which includes all of the QSOs in the HST archives (available as of 2001) matching these redshift, magnitude, and instrumental criteria.  The overall sample is chosen to probe a broad range of QSO properties, rather than focussing on particular classes of QSO.  The subset of 42 used here are those which also have nuclear x-ray data available in the literature and which have either an elliptical host or a spiral with a bulge that can be modeled.

The redshift range is chosen to allow the separation of the nuclear point spread function (PSF) from the host, which is performed with a non-linear least squares fit to the 2-D image.  The fitting routine first uses an oversampled artificial PSF generated by the {\sc tinytim} software package to model focus and subpixel centring.  Secondly, it performs a simultaneous fit to the PSF + host.  The host is modeled with either an exponential disc (spiral) or an $r^{1/4}$ law (ellipticals and elliptical bulges).  Details of the modeling are given by Hamilton et al.~(2002).  Bulges of spiral hosts are modeled separately from the discs.  From the model, we are able to measure $r_e$, the half-light radius, $\mu_e$, the effective surface magnitude (surface magnitude at the half-light radius), and the $V$-band apparent magnitude.  Magnitudes are K-corrected and converted into rest-frame $V$-band absolute magnitudes.

Radio-loudness and x-ray luminosity data are collected from the literature.  Radio-loudness, mostly taken from Brinkmann, Yuan \& Siebert~(1997) and Yuan et al.~(1998), is defined as having a radio-to-optical flux density ratio greater than 10.  Nuclear x-ray luminosities, $L_X$ (erg s$^{-1}$), are taken from the literature and mostly come from {\it ROSAT} and {\it Einstein} observations.  They are normalized to $\nu L_\nu$ at a rest-frame energy of 0.5 keV.  A complete list of literature sources is given by Hamilton et al.~(2002).  The sample includes 20 radio loud quasars, 22 radio quiet quasars, 33 elliptical hosts, and 9 spirals.

A Principal Components Analysis is performed on the host properties ($\mu_e, \log r_e$) and the nuclear luminosity ($\log L_X$).  The projections of these principal components  (or eigenvectors) in $\{\log L_X, \mu_e, \log r_e\}$ space are given in Table~\ref{table:pc-projections}.  The PCA was performed using IDL's {\sc pca} procedure in the {\sc astrolib} library.  Note that its scaling of the principal component projections is different than in IDL's standard {\sc pcomp} procedure.  The projections are made on to the space of normalized variables (of zero mean and unit variance) and are not scaled to the eigenvalues.

\begin{figure*}
\psfig{figure=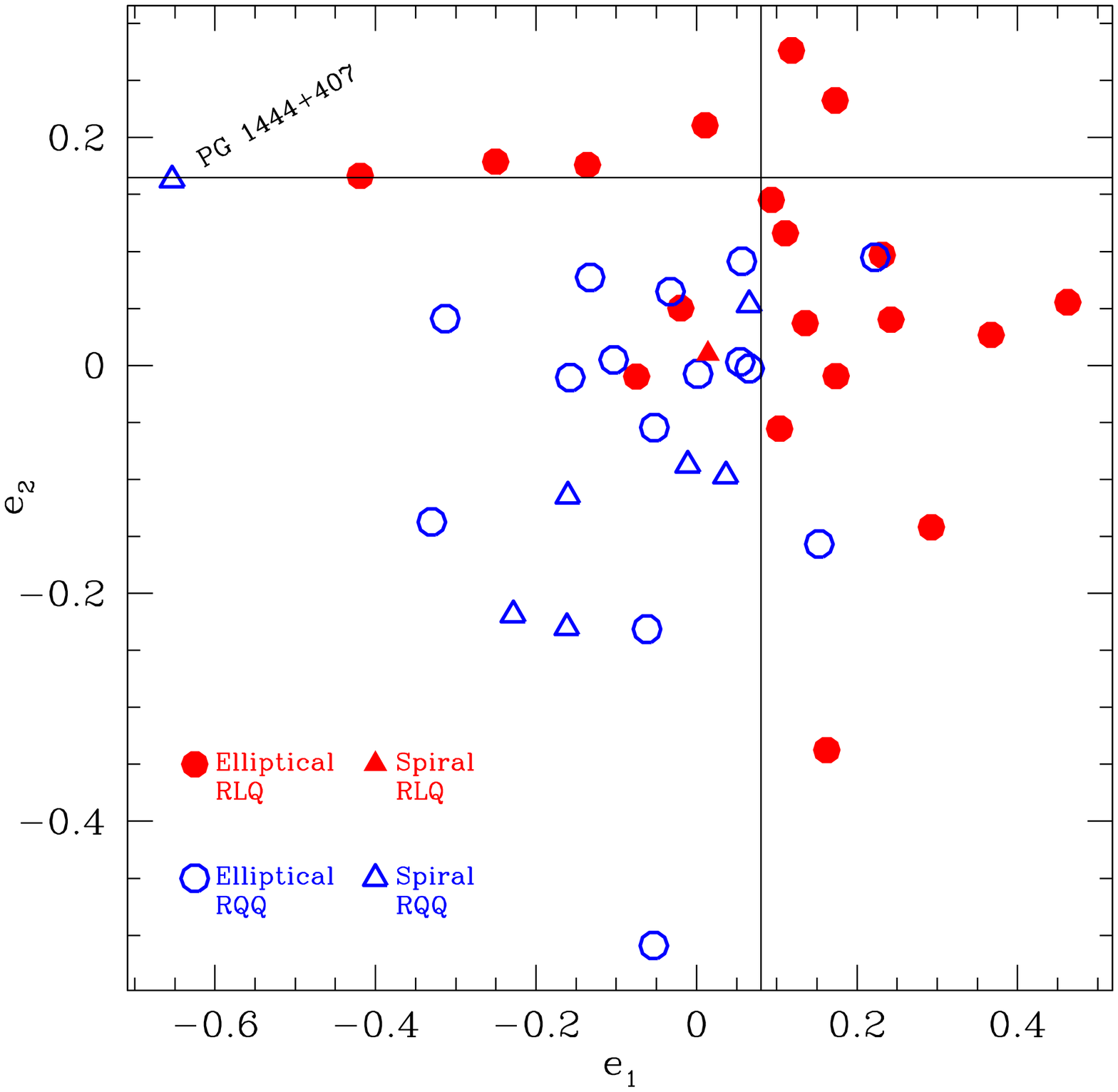,width=8cm,angle=0}
\caption{Distribution of objects in the QSO fundamental plane space, formed by the first ($\bmath{e_1}$) and second ($\bmath{e_2}$) principal components of the sample.  Radio-louds are solid figures, radio-quiets are hollow, elliptical hosts are circles, and spirals are triangles.  RLQs are primarily found above $\bmath{e_1} \ga 0.08$ or $\bmath{e_2} \ga 0.165$.  RQQs are inside the area below both these limits.  Note that the spirals are confined to a narrow, diagonal strip near the centre, with the exception of PG~1444+407.} 
\label{fig:xfp}
\end{figure*}

The first principal component, $\bmath{e_1}$, is dominated by the host properties and is qualitatively similar to the Kormendy relation (Kormendy~1977; Hamabe \& Kormendy~1987), a correlation between $\log r_e$ and $\mu_e$.  The second principal component, $\bmath{e_2}$, is dominated by the nuclear x-ray luminosity.  The first two principal components define a QSO `fundamental plane', which can be expressed as $\log L_X = 79.3 - 2.03 \mu_e + 8.74 \log r_e$, giving a relationship between the host and nuclear properties.  (The final principal component, representing only about 5 per cent of the sample variance, is the plane's thickness.)  But if we plot the QSOs in the $\{\bmath{e_1}, \bmath{e_2}\}$ plane itself (Fig.~\ref{fig:xfp}), we can look for groupings of the QSO types and use the principal components as a classification tool.  We find that different QSO types lie in different regions of the fundamental plane.  

The most obvious clustering is in radio-loudness.  Radio-loud QSOs (RLQ) dominate the regions at large $\bmath{e_1}$ or large $\bmath{e_2}$, while radio-quiet QSOs (RQQ) dominate the region with low $\bmath{e_1}$ and $\bmath{e_2}$.  Is this clustering real, or are both classes drawn from the same distribution in the fundamental plane?  We can form a null hypothesis that the RLQs and RQQs are drawn from the same parent population.  A two-dimensional Kolmogorov--Smirnov (K--S) test (Peacock 1983; Fasano \& Franceschini 1987) returns a statistic of $D=0.675$, where the probability of obtaining $D>0.675$ under the null hypothesis is $P(>D) \sim 0.0003$.  Thus, at the 5 per cent significance level, we have sufficient evidence to conclude that the radio-loud and radio-quiet quasars are not drawn from the same distribution.

RLQs dominate the regions above $\bmath{e_1} \ga 0.08$ or $\bmath{e_2} \ga 0.165$, while RQQs are found in the zone below both these limits.  The dividing lines are found empirically by minimizing the total contamination rate between the groups, one dimension at a time.  There are two groups (radio-loud and radio-quiet) and two zones along an axis (either side of the dividing line).  With our data set, we can divide the space so that in one zone, the majority of QSOs are RLQs, and a minority are RQQs (the opposite is true for the other zone).  The contamination rate of the zone is the fraction of its objects that are in the minority.  To find the best dividing line for one axis, we minimize the sum of the contamination rates of both zones.  Each axis is treated independently.  Given the spacing of the QSOs, the division along $\bmath{e_1}$ can be anywhere between 0.0660 and 0.0933, and the $\bmath{e_2}$ division can be between 0.1626 and 0.1664.  With these limits, there are five (out of 42) QSOs that are `off-sides'.  Out of 20 QSOs in the radio-loud zone, two are RQQs, a 10.0 per cent contamination.  And out of 22 in the radio-quiet zone, three are RLQs, which is a 13.6 per cent contamination.

Clustering by host morphology shows that the bulges of most spiral hosts occupy a very restricted region--a narrow diagonal running from near the origin down to $(-0.2, -0.2)$.  Our one radio-loud spiral (3C~351) is within this zone, and in fact it appears to be not only typical for a spiral bulge, but it is almost perfectly `average' for the galaxies in general, lying near the weighted centre of the distribution at $(0,0)$.  A notable exception to the spirals is PG~1444+407, which is an extreme outlier.  Its unusually compact bulge is very elliptical (ellipticity, $\epsilon=0.43$; see Hamilton~2001) and may in fact be a bar, as speculated by Bahcall et al.~(1997).  Thus its placement on the far left of Fig.~\ref{fig:xfp} could be a matter of using an inappropriate model.  Elliptical hosts, in contrast to spirals, are found over the entire range of the plot, including the zone occupied by the spirals.  If we include PG~1444+407 in the sample, a two-dimensional K--S test on the distribution of morphologies returns a statistic of $D=0.525$, with $P(>D) \sim 0.009$, while if we exclude it, we find $D=0.575$ and $P(>D) \sim 0.003$.  Note how much more pronounced the difference in the distributions is if we reject PG~1444+407 as incorrectly modeled.  In either case, we again have enough evidence at the 5 per cent significance level to conclude that the QSOs in elliptical and spiral hosts are not drawn from the same distribution.

The clustering within the fundamental plane is interesting, but it is still something of an abstraction at this point.  We can draw more useful conclusions about the radio-loudness distribution if we interpret the principal axes physically.

\section{Kormendy Relation}\label{sec:kormendy}

\subsection{Size Distribution}\label{sec:kormendy-size}

\begin{figure*}
\psfig{figure=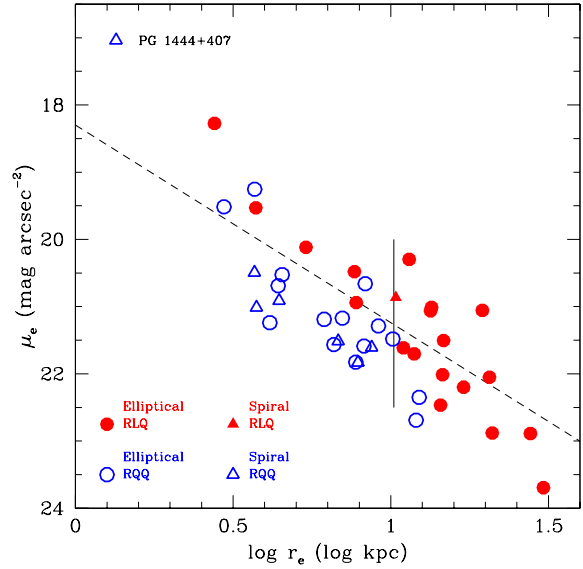,width=8cm,angle=0}
\caption{Distribution of the complete sample of QSOs in $\mu_\mathrm{e}$ and $r_\mathrm{e}$.  Radio-louds are solid figures, radio-quiets are hollow, elliptical hosts are circles, and spirals are triangles.  The vertical line divides our large and small hosts, and the Kormendy relation is shown by the dashed line.  PG~1444+407 lies away from the Kormendy relation trend and is again (as in Fig.~\ref{fig:xfp}) an outlier with respect to the spiral bulges.}
\label{fig:kormendy}
\end{figure*}

As stated above, $\bmath{e_1}$ is similar to the Kormendy relation, and $\bmath{e_2}$ is dominated by x-ray luminosity, so we turn next to physical properties.  To show the Kormendy relation, we plot the QSOs in $\{ r_e, \mu_e \}$ space, Fig.~\ref{fig:kormendy}.  The RLQs and RQQs are distributed differently in size.  The greatest separation in their cumulative distributions is found at $r_e=10.2$ kpc, marked by the vertical line.  The large hosts ($r_e \geq 10.2$ kpc) are dominated by RLQs, and the small hosts are dominated by RQQs.  The significance of this division will be explored in more detail in \S\ref{sec:association}.  

\subsection{Luminosity Distribution}\label{sec:kormendy-luminosity}

\begin{figure*}
\begin{tabular}{cc}
\psfig{figure=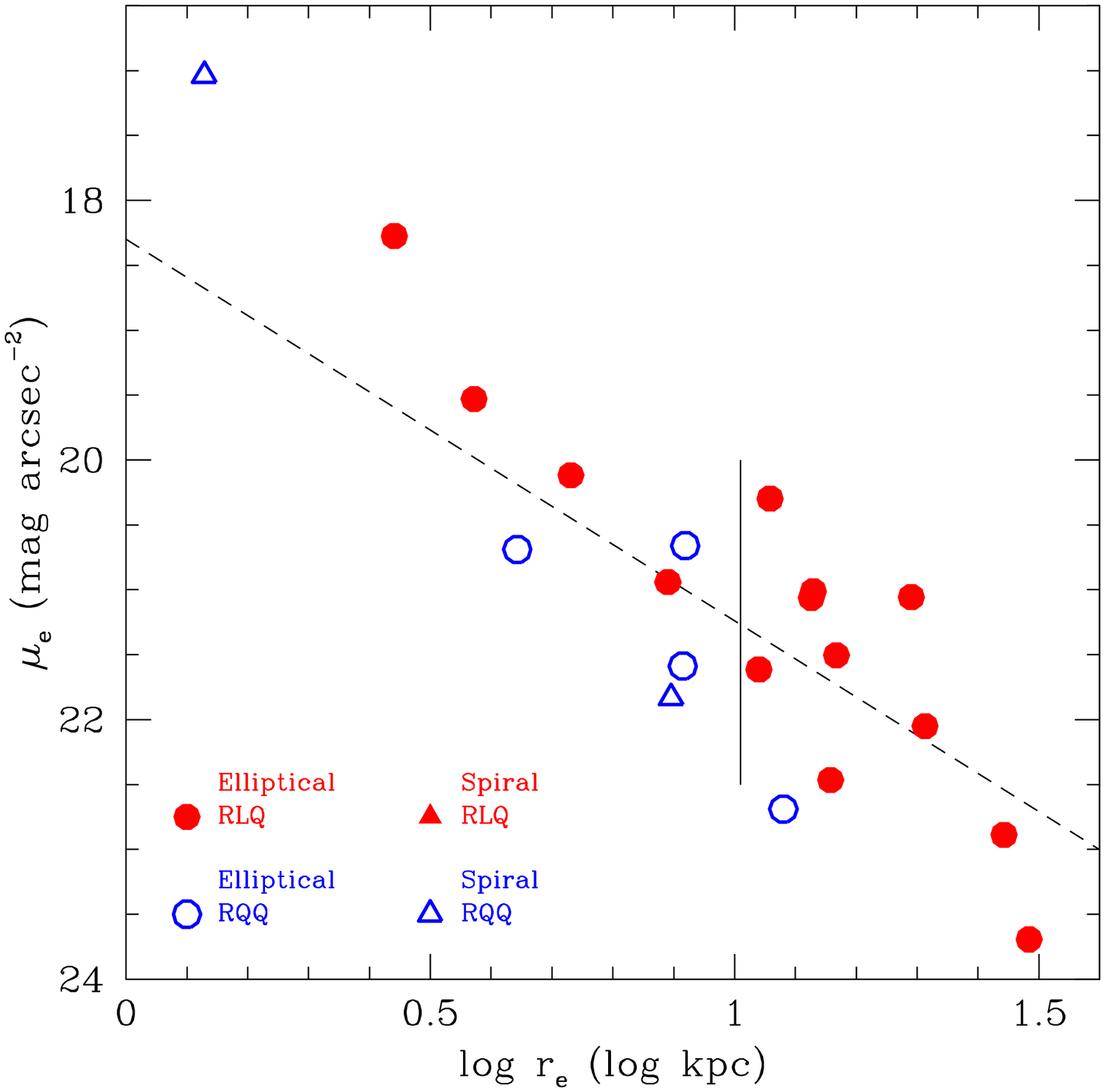,width=8cm,angle=0}&
\psfig{figure=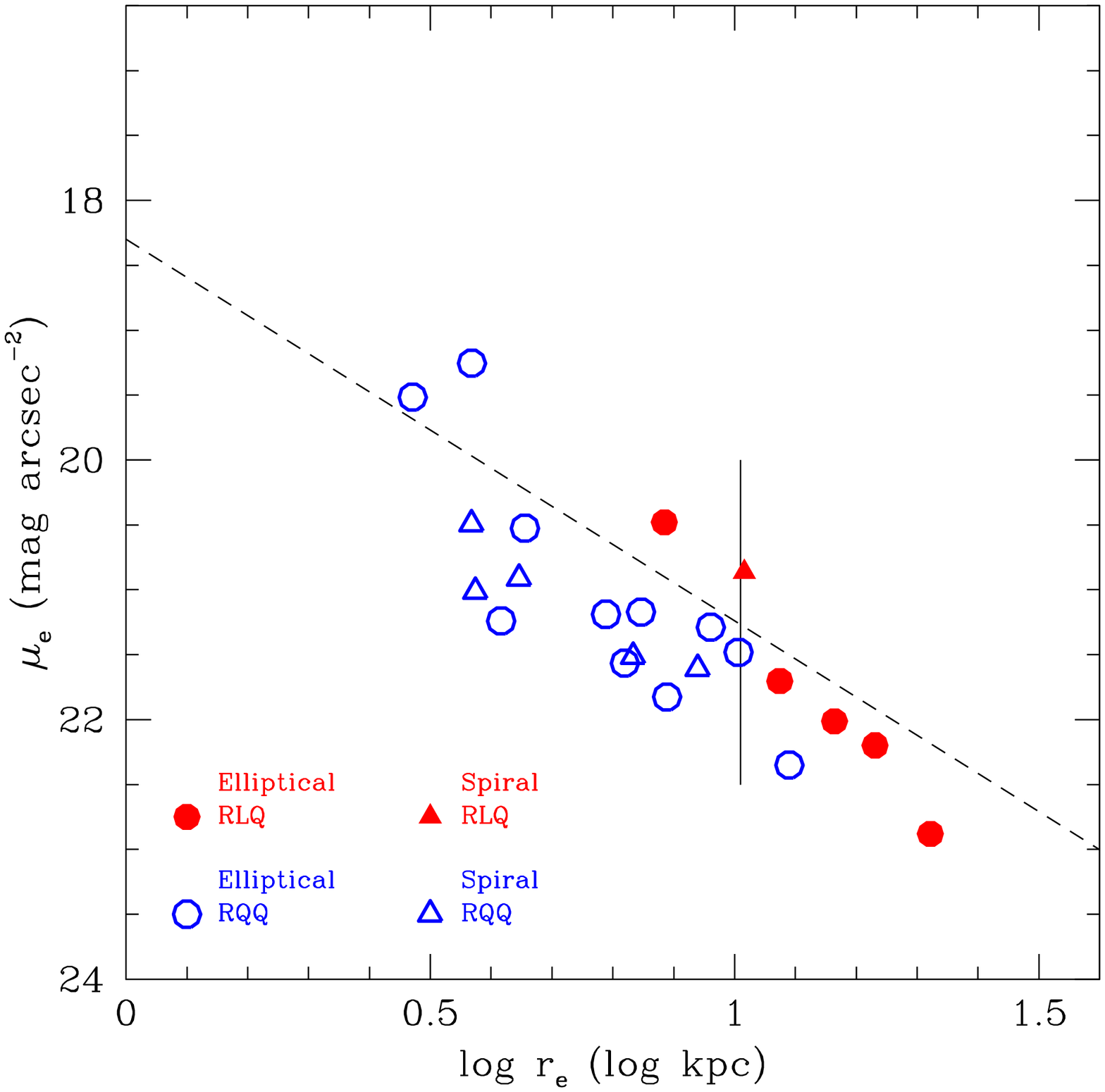,width=8cm,angle=0}\\
\end{tabular}
\caption{Distribution of QSOs in $\mu_\mathrm{e}$ and $r_\mathrm{e}$, separated by nuclear luminosity.  The left panel shows high-luminosity QSOs ($L_X \geq 10^{44.5}$ erg s$^{-1}$), and the right panel shows low-luminosity QSOs.  Radio-louds are solid figures, radio-quiets are hollow, elliptical hosts are circles, and spirals are triangles.  The vertical line divides our large and small hosts (which we believe to be dominated by core and coreless galaxies, respectively), and the Kormendy relation is shown by the dashed line.  The high-luminosity QSOs are nearly all radio-loud, but at low luminosities, there is a tendency to divide into RLQ and RQQ classes, according to whether the host is core or coreless, respectively.}
\label{fig:kormendy-split}
\end{figure*}

In x-ray luminosity, the greatest separation in the cumulative distributions of RLQs and RQQs is at $L_X = 10^{44.5}$ erg s$^{-1}$.  If we next look at just the QSOs with $L_X \ge 10^{44.5}$ erg s$^{-1}$, shown in the left panel of Fig.~\ref{fig:kormendy-split}, we find that these high-luminosity objects are dominated by RLQs (14 out of 20).  The high-luminosity RLQs span nearly the full sample's range of sizes, from the largest object (30.5 kpc) down to the second-smallest (2.8 kpc).

On the other hand, when we look at the lower-luminosity objects (the right panel of Fig.~\ref{fig:kormendy-split}), we see they are dominated by RQQs (16 out of 22).  Furthermore, there is a fairly clean division in the sizes of RLQs and RQQs.  Looking again at our 10.2 kpc dividing line, all but one of the RLQs are in `large' hosts, and all but one RQQs are in `small' hosts.

\subsection{Redshift Cuts}\label{sec:kormendy-z}

\begin{figure*}
\begin{tabular}{cc}
\psfig{figure=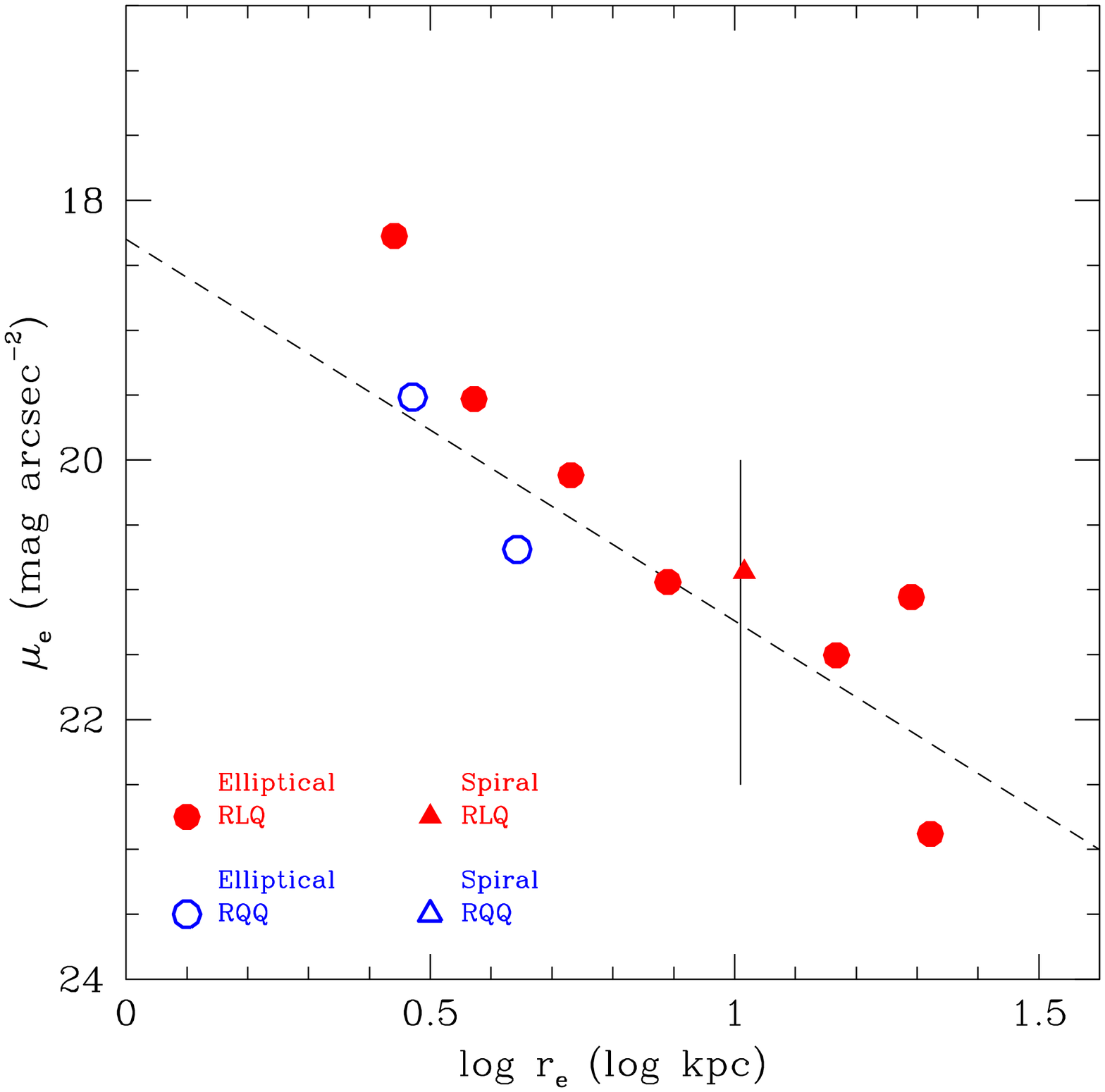,width=8cm,angle=0}&
\psfig{figure=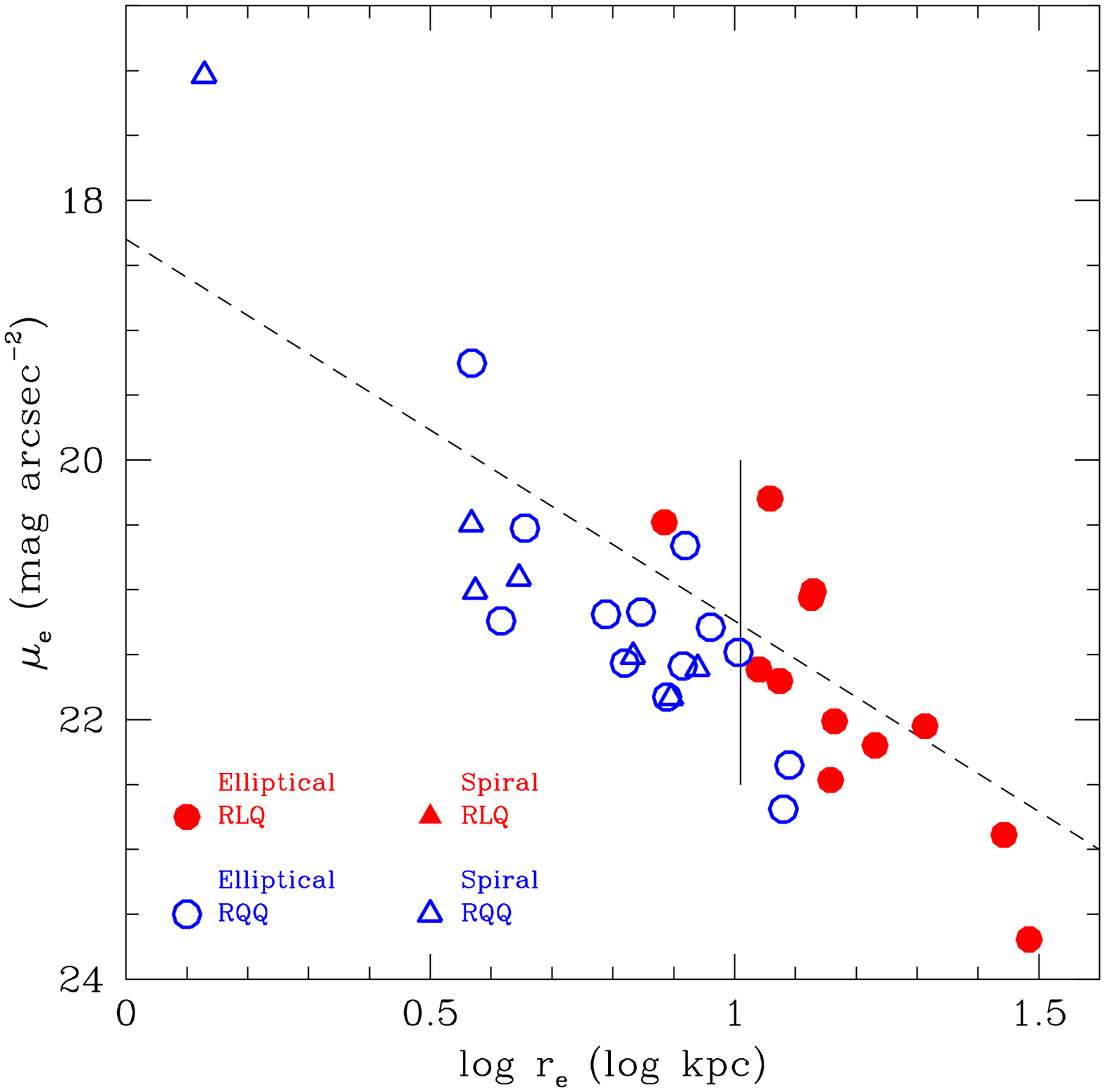,width=8cm,angle=0}\\
\end{tabular}
\caption{Distribution of QSOs in $\mu_\mathrm{e}$ and $r_\mathrm{e}$, separated by redshift.  Higher-redshift objects ($z \ge 0.35$) are on the left, and low-redshift objects ($z < 0.35$) are on the right.  Radio-louds are solid figures, radio-quiets are hollow, elliptical hosts are circles, and spirals are triangles.  The vertical line divides our large and small hosts (which we believe to be dominated by core and coreless galaxies, respectively), and the Kormendy relation is shown by the dashed line.  At higher redshift, RLQs are found in both core and coreless types, but the coreless RLQs are mostly lost by $z<0.35$.}
\label{fig:kormendy-z-split}
\end{figure*}

We can make a cut in redshift to look for biases and evolutionary effects (Fig.~\ref{fig:kormendy-z-split}).  When we look at the Kormendy relation distributions of radio-loud and radio-quiet QSOs, we find that below $z < 0.35$, the behaviour of RQQs and RLQs is similar to that found for the low-luminosity sample, with all but two of the 20 RQQs having small hosts and all but one of the 12 RLQs having large hosts.  For $z \ge 0.35$, the RLQs act much more like the high-luminosity sample, being spread across nearly the full range of sizes and populating both the large and small classes.

\section{Association with Core and Coreless Ellipticals}\label{sec:association}

Earlier work by McLure et al.~(1999) has already shown that RLQs have more luminous hosts than RQQs.  In our study, we show that they also tend to be larger.  As described above, we have found the greatest difference between the RLQ and RQQ cumulative size distributions to be at $r_e = 10.2$ kpc.  In the study of Kormendy et al.~(2009), core and coreless elliptical galaxies are divided near the same point, with core ellipticals having $r_e \ga 10$ kpc, while coreless ellipticals and the bulges of disc galaxies are smaller.  Thus it seems likely that the RLQs occur primarily in core ellipticals, while the RQQs occur primarily in coreless ellipticals and spiral bulges, and that these associations are much more pronounced at low x-ray luminosities and at low redshifts. 

Kormendy et al.~(2009) describe core ellipticals as being brighter, with boxy isophotes, generally older stars, and x-ray emitting gas.  They are often associated with a strong nuclear radio source.  Coreless ellipticals, on the other hand, are fainter, with discy isophotes and usually without x-ray emitting gas.  They rarely have strong radio sources.  It is theorized that core ellipticals form from the dissipationless `dry' mergers of multiple galaxies at once.  The coreless ellipticals are thought to form in a history of fewer and dissipative `wet' galaxy mergers.

All of these properties are known for inactive galaxies and non-QSOs generally.  Assuming we can make a general association between host size and the presence of a core, then the data presented in our study show that for low-redshift or low-luminosity QSOs, radio-loudness is associated with core elliptical hosts, and radio-quietness with coreless ellipticals and spirals.  High-redshift and high-luminosity QSOs in general are likely to be radio-loud, regardless of whether they have core or coreless hosts.  In the remainder of our analysis, we assume the larger hosts are more likely to be core, and the smaller ones are more likely to be coreless, although there may not be a clean division at $r_e = 10.2$ kpc.

\section{Discussion}\label{sec:discussion}

Quasars cluster in the QSO fundamental plane both by morphology and, more cleanly, radio-loudness, permitting this plane to be used as a classification tool.  An interesting comparison can be made with the 4D Eigenvector 1 space described by Zamfir, Sulentic \& Marziani~(2008), in which there is also a separation of radio-loud and radio-quiet populations.  
Given a QSO's location on the QSO fundamental plane, we can correctly predict its radio-loudness 88.1 per cent of the time.  Part of the clustering in the plane appears to be connected to distinctions between core and coreless galaxies.  Among non-QSOs and our low luminosity QSO sample, strong radio emission is associated with core but rarely with coreless ellipticals.  The distinction only seems to be broken in the highest-luminosity QSOs.  

\subsection{Luminosity and Accretion}\label{sec:discussion-accretion}

Does accretion rate play a role in breaking this apparent dichotomy?  Since x-ray luminosity is generally associated with accretion rate, then maybe what we are seeing is that a high accretion rate can overcome the conditions that prevent strong radio emission in coreless hosts (though this picture is complicated by x-ray emission from jets).  Thick-disc, radiatively-inefficient accretion flows are often thought to be associated with radio jets, and they can exist at either high or low accretion rates.  Perhaps the tripartite division of quasars into low-luminosity RQQ, low-luminosity RLQ, and high-luminosity RLQ reflects a division of accretion flows between thin discs, thick discs with low accretion rate, and thick discs with high accretion rate.  

Since the core/coreless division among ellipticals is taken as an indicator of merger history, then this would set up an empirical connection between the accretion type and the galaxy merger history, as well.  But because of the complexities of the various accretion models, any simple statement is difficult to make at this point.  In particular, the issue of how accretion flow in the central engine is affected by galactic merger history appears to be an open question (for some existing work, see Liu~2004).  More observations and modeling on this front are needed.  Furthermore, we must also keep in mind that the RQQ classification used here could be complicated by the duty cycle of the radio sources.

\subsection{Redshift and Evolution}\label{sec:discussion-evolution}

We see two trends as we go from high to low redshift.  First, the coreless RQQs become a larger fraction of the QSO population.  And perhaps more significantly, among RLQs there is a loss of coreless hosts.  Are these trends a matter of redshift bias?  We are more likely to miss faint objects at higher $z$, and the RQQs are in fainter hosts, so the increase in coreless RQQs may have a bias component.  The original selection methods of these QSOs include a mix of criteria (x-ray selected, stellar appearance, etc.; see Hamilton et al.~2008 for a detailed analysis), but only 2 of the 18 low-$z$ small ($r_e < 10.2$ kpc) RQQs have fainter hosts (in total host magnitude) than the faintest of the high-$z$ small RQQ set.

The loss of coreless RLQs is a different matter.  These are seen at high redshift, where they are more difficult to detect, but are not seen at low redshift.  Furthermore, these occur in bright hosts that wouldn't be missed at low $z$.  This trend seems less susceptible to a detection bias, and we appear to be seeing evolutionary effects in the loss of coreless RLQs.  Perhaps over time, the most massive coreless hosts have undergone major mergers that turn them into core ellipticals, leaving only the least massive coreless hosts, which contain radio-quiet QSOs.

In this short study, we have not established whether nuclear luminosity or evolution is the primary factor driving the segregation of RLQs into core and RQQs into coreless hosts.  Effects such as the downsizing of QSOs entangle the two, but this will be left for a later study.

\section{Acknowledgments}

The author would especially like to thank Martin Gaskell of the University of Texas, for pointing out the possible connection between the QSO fundamental plane results and the core/coreless galaxy division.  Thanks to Dirk Grupe of Penn State, for comments on x-ray emission and accretion models, to Massimo Stiavelli of STScI, for his advice on galaxies and statistics, and to Jeffrey Newman of the University of Pittsburgh, for his recommendation to look at redshift.  Thanks also to Doug Darbro and Woonyuen Koh of Shawnee State University, who helped with the statistical tests.

{}


\begin{table*}
\centering
\begin{minipage}{140mm}
  \caption{PCA Results}
  \begin{tabular}{cccccc}
  \hline
  $\begin{array}[t]{c} \mbox{Principal}\\ \mbox{component} \end{array}$ & $\begin{array}[t]{c} \lambda \\ \mbox{(per cent)} \end{array}$ & $\begin{array}[t]{c} \mbox{Cumulative} \\ \mbox{percentage} \end{array}$  & $\begin{array}[t]{c} \mbox{Projection on to} \\ \log L_X  \end{array}$ & $\begin{array}[t]{c} \mbox{Projection on to} \\ \mu_e  \end{array}$ & $\begin{array}[t]{c} \mbox{Projection on to} \\ \log r_e  \end{array}$ \\
  \hline
$\bmath{e_1}$ & 61.7 &  61.7 & 0.16 & 0.69 & 0.71 \\
$\bmath{e_2}$ & 33.5 &  95.2 & 0.97 & -0.24 & 0.01 \\
$\bmath{e_3}$ &  4.8 & 100.0 &  &  &  \\
\hline
\end{tabular}
\label{table:pc-projections}
\medskip
Col.~(2), eigenvalue as percentage of total variance. Col.~(3), cumulative percentages.  Cols.~(4-6), projections of principal components on to normalized variables, not scaled for eigenvalues.  The third principal component's projections are not shown, since it is not included in the fundamental plane.\end{minipage}
\end{table*}


\begin{thebibliography}{}

\bibitem[Bahcall et al.(1997)]{Bahcall97} Bahcall J.  N., Kirhakos S., Saxe D.  H., 1997, ApJ, 479, 642

\bibitem[Boroson \& Green(1992)]{Boroson92} Boroson T. A., Green R. F., 1992, ApJS, 80, 109

\bibitem[Brinkmann et al.(1997)]{Brinkmann97} Brinkmann W., Yuan W., 
Siebert J., 1997, A\&A, 319, 413

\bibitem[Crane et al.(1993)]{Crane93} Crane P., Stiavelli M., King I. R. et al., 1993, AJ, 106, 1371

\bibitem[Diehl \& Statler(2005)]{Diehl05} Diehl S., Statler T. S., 2005, ApJ, 633, L21

\bibitem[Djorgovski \& Davis(1987]{Djorgovski87}  Djorgovski S., Davis M., 1987, ApJ, 313, 59 

\bibitem[Dressler et al.(1987)]{Dressler87} Dressler A., Lynden-Bell D., Burstein D., Davies R. L., Faber S. M., Terlevich R. J., 
Wegner G., 1987, ApJ, 313, 42 

\bibitem[Falcke et al.(1996)]{Falcke96} Falcke H., Sherwood W., Patnaik A. R., 1996, ApJ, 471, 106

\bibitem[Fasano \& Franceschini(1987)]{Fasano87} Fasano G., Franceschini A., 1987, MNRAS, 225, 155

\bibitem[Ferrarese \& Merritt(2000)]{Ferrarese00} Ferrarese L., Merritt D., 2000, ApJ, 539, L9

\bibitem[Gebhardt et al.(2000)]{Gebhardt00} Gebhardt K., Bender R., Bower G. et al., 2000, ApJ, 539, L13

\bibitem[Guyon et al.(2006)]{Guyon06}  Guyon O., Sanders D. B., Stockton A., 2006, ApJS, 166, 89

\bibitem[Hamabe \& Kormendy(1987))]{Hamabe87} Hamabe M.,  Kormendy J., 1987, in Structure and Dynamics of
Elliptical Galaxies, ed. T. de Zeeuw (Dordrecht, Holland: IAU), 379

\bibitem[Hamilton(2001)]{Hamilton01} Hamilton T. S., 2001, PhD thesis, Univ. of Pittsburgh

\bibitem[Hamilton et al.(2002)]{Hamilton02} Hamilton T. S., Casertano S.,  Turnshek D. A., 2002, ApJ, 576, 61

\bibitem[Hamilton et al.(2008)]{Hamilton08} Hamilton T. S., Casertano S.,  Turnshek D. A., 2008, ApJ, 678, 22

\bibitem[Kormendy(1977)]{Kormendy77} Kormendy J., 1977, ApJ, 217, 406

\bibitem[Kormendy et al.(2009)]{Kormendy09} Kormendy J., Fisher D. B., Cornell M. E., Bender R., 2009, ApJS, 182, 216

\bibitem[Laor \& Behar(2008)]{Laor08} Laor A., Behar E., 2008, MNRAS, 390, 847

\bibitem[Liu(2004)]{Liu04} Liu F. K., 2004, MNRAS, 347, 1357

\bibitem[McLure et al.(1999)]{McLure99} McLure R. J., Kukula M. J., Dunlop J. S., Baum S. A., OÕDea C. P., Hughes D. H., 1999, MNRAS, 308, 377

\bibitem[Magorrian et al.(1998)]{Magorrian98} Magorrian J., Tremaine S., Richstone D. et al., 1998, AJ, 115, 2285

\bibitem[Merloni et al.(2003)]{Merloni03} Merloni A., Heinz S., di Matteo T., 2003, MNRAS, 345, 1057

\bibitem[Peacock(1983)]{Peacock83} Peacock J. A., 1983, MNRAS, 202, 615

\bibitem[Trager et al.(2000)]{Trager00}  Trager S. C., Faber S. M., Worthey G., Gonz\'{a}lez J. J., 2000, AJ, 120, 165

\bibitem[Trujillo et al.(2004)]{Trujillo04} Trujillo I., Erwin P., Ramos A. A., Graham A. W., 2004, AJ, 127, 1917

\bibitem[Tsutsui et al.(2009)]{Tsutsui09} Tsutsui R., Nakamura T., Yonetoku D., Murakami T., Kodama Y., Takahashi K., 2009, J. Cosmology and Astroparticle Phys., 8, 15

\bibitem[Wang et al.(2003)]{Wang03} Wang J.-M., Ho L. C., Staubert R., 2003, A\&A, 409, 887

\bibitem[Whitmore(1984)]{Whitmore84} Whitmore B. C., 1984, ApJ, 278, 61

\bibitem[Yip et al.(2004)]{Yip04} Yip C. W., Connolly A. J., Vanden Berk D. E. et al., 2004, AJ, 128, 2603

\bibitem[Yuan et al.(1998)]{Yuan98} Yuan W., Brinkmann W., Siebert
J., Voges W., 1998, A\&A, 330, 108

\bibitem[Zamfir et al.(2008)]{Zamfir08} Zamfir S., Sulentic J. W., Marziani P., 2008, MNRAS, 387, 856

\end{thebibliography}
\end{document}